\documentclass[a4paper,12pt]{article}

\usepackage{amssymb}
\usepackage{amsmath}
\usepackage{bbm}
\usepackage{mathbbol}
\usepackage{color}
\usepackage{graphicx}
\usepackage{mathrsfs}
\usepackage{pgfplots}
\usepackage{tikz}
\usetikzlibrary{patterns}
\usepackage{pst-plot,pst-node}
\usepackage[all]{xy}
\usepackage{kotex}

\newtheorem{theorem}{Theorem}
\newtheorem{lemma}{Lemma}
\newtheorem{proposition}{Proposition}
\newtheorem{definition}{Definition}

\newcommand\unit{\hbox{\rm 1\kern-2.8truept l}}

\newcommand\mi{\mathrm{i}}

\newcommand\Dom{\hbox{\rm Dom}}

\newcommand\Lform{{\mathcal{L}}\kern-7.56pt\raise1.0pt\hbox{$-$}}

\font\rm=cmr12

\font\rmseven=cmr8

\begin{document}

\title{The Kossakowski Matrix and Strict Positivity of Markovian Quantum Dynamics}
\author{Juli\'an Agredo, Franco Fagnola and Damiano Poletti}
\date{}

\maketitle
\abstract{We investigate the relationship between strict positivity of the Kossakowski matrix,
irreducibility and positivity improvement properties of Markovian Quantum Dynamics.
We show that for a Gaussian quantum dynamical semigroup strict positivity of the Kossakowski
matrix implies irreducibility and, with an additional technical assumption, that the support
of any initial state is the whole space for any positive time.}




\section{Introduction}
\setcounter{equation}{0}

The Kossakowski matrix was introduced in \cite{GKS} (Theorem 2.2, see \cite{ChPa} for a detailed account)
to represent the generator of a completely positive semigroup on the algebra $M_n(\mathbb{C})$ of $n\times n$
complex matrices with respect to a basis of traceless matrices
\[
\rho \mapsto -\mi [H,\rho] +\frac{1}{2}\sum_{j,k=1}^{d^2-1}c_{kj}\left([F_k,\rho\,F_j^*]+[F_k\rho,F^*_j]\right)
\]
where $H=H^*\in M_d(\mathbb{C})$, $F_k\in M_n(\mathbb{C})$, $\hbox{\rm tr}(F_k)=0$, $\hbox{\rm tr}(F_k F^*_j)=\delta_{kj}$.
The hermitian matrix $[c_{kj}]_{1\leq k,j\leq d^2-1}$ is now called the \emph{Kossakowski matrix}.
Matrices $F_k$, together with the identity matrix, form a basis of $M_n(\mathbb{C})$. In the Heisenberg picture the
generator is written
\begin{equation}\label{eq:GKLSbd}
\mathcal{L}(x) = \mi [H,x] + \frac{1}{2}\sum_{j,k=1}^{d^2-1}c_{kj}\left([F^*_j,x]F_k+F^*_j[x,F_k]\right)
\end{equation}

A similar concept had already been considered by Andrzej Kossakowski in \cite{Koss} (Theorem 11) to
write a GKLS representation of the generator of classical Brownian motion on an $n$-dimensional Lie group
to quantify noises appearing in the Markovian quantum master equation. It also emerges in the
representation of generators of Gaussian quantum dynamical semigroups (see \cite{Po2021} and the
references therein) but the bounded operators $F_j$ are replaced by unbounded creation and
annihilation operators therefore the GKSL representation appears in generalized form and must
be handled with more care (see \cite{AFP-MJM} for details).

In this paper we investigate consequences of strict positivity of the Kossakowski matrix (Definition \ref{def:Koss-mat})
of a Gaussian quantum Markov semigroup (QMS) (defined in Sect. \ref{sect:Koss-mat}) on irreducibility and strict positivity
of the corresponding Markovian dynamics. Gaussian QMSs arise in several relevant models and form a class with a rich structure
with a number of explicit formulas (see e.g. \cite{Te} and the references therein).

Following the terminology in use in the classical theory of Markov processes we recall the following
\begin{definition}\label{def:sh-ir-pi}
Let $\mathcal{T}$ be a QMS.
\begin{enumerate}
\item A projection $p$ is \emph{subharmonic} if  $\mathcal{T}_t(p)\geq p$ for all $t\ge 0$.
\item $\mathcal{T}$ is \emph{irreducible} if its only subharmonic projections are $0$ and $\unit$.
\item $\mathcal{T}$ is \emph{positivity improving} if $\mathcal{T}_t(x)>0$ for all non-zero $x\geq 0$ and $t>0$.
\end{enumerate}
\end{definition}

Irreducibility is an important useful property in the analysis of the dynamics because it allows one to establish
from the outset that the system has to be regarded as a whole and reduction to invariant subsystems is not possible.
In particular, the support of any initial state cannot remain confined in a proper subspace and, as a consequence,
it can be looked as a weak reachability condition as in quantum control. Theorem \ref{th:irreduc} shows that strict
positivity of the Kossakowski matrix implies irreducibility.

Property 3, also called immediate positivity, follows from L\'evy-Austin-Ornstein theorems for certain classical Markov processes.
Extending these results it has been proved in \cite{GlHaNa} for a class of quantum Markovian dynamics.
Theorem \ref{th:pos-imp} proves that strict positivity of the Kossakovski matrix implies that a QMS with
strictly positive Kossakowski matrix is positivity improving, if a certain semigroup is analytic. This
assumption is needed to deal with unboundedness of the generator.

The paper is organized as follows: in Sect.~\ref{afpsect:fdcase} we present the finite dimensional
version of our results. In Sect.~\ref{sect:Koss-mat} we introduce the Kossakowki matrix of a Gaussian
QMS and show that its strict positivity implies irreducibility. Positivity improvement is considered
in Sect.~\ref{sect:pos-impr} after some preliminary results on domains of the unbounded operators
arising in our framework. We also present an example with some indication on how one can check that
a certain operator generates an analytic semigroup.
Final conclusions are collected in the last section.

\section{The finite dimensional case}\label{afpsect:fdcase}

In this Section we show that if the Kossakowski matrix has full rank then the QMS is positivity
improving whence irreducible. This fact is well-known (see e.g. \cite{BPT}) but we give another proof
to ease the path to the infinite dimensional situation.

\begin{proposition}
If the Kossakowski matrix is strictly positive, then the QMS generated by (\ref{eq:GKLSbd}) is positivity improving.
In particular, it is irreducible.
\end{proposition}

\noindent{\it Proof.} Note that, for all $u,v\in\mathsf{h}$ with $\Vert u\Vert=\Vert v\Vert=1$ the map
$t\mapsto \left\langle v,\mathcal{T}_t(|u\rangle\langle u|)v\right\rangle$
is strictly positive on an interval $]0,t_{u,v}[$ with $t_{u,v}>0$.

Indeed, this is clear if $v$ is not orthogonal to $u$ because
$\left\langle v,\mathcal{T}_t(|u\rangle\langle u|)v\right\rangle>0$ in a right neighbourhood of $0$
by time continuity.
If $\langle v,u\rangle=0$ let $G=-\mi H -(1/2)\sum_{jk}c_{jk}F^*_jF_k$,
let $P_t=\hbox{\rm e}^{tG}$ and note that
\[
\frac{\hbox{\rm d}}{\hbox{\rm d}s} \mathcal{T}_s(|P_{t-s}^*u\rangle\langle P_{t-s}^*u|)
= \sum_{j,k=1}^{d^2-1}c_{jk}\mathcal{T}_s(|F_j^*P_{t-s}^*u\rangle\langle F_k^*P_{t-s}^*u|)
\]
therefore, taking the scalar product with $v$ and integrating on $[0,t]$
\[
\left\langle v,\mathcal{T}_t(|u\rangle\langle u|)v\right\rangle
= \left|\langle v,P_t^* u\rangle\right|^2
+  \sum_{j,k=1}^{d^2-1}c_{jk}\int_{0}^t
\left\langle v,\mathcal{T}_s(|F_j^*P_{t-s}^*u\rangle\langle F_k^*P_{t-s}^*u|)v\right\rangle \hbox{\rm d}s.
\]
If $\langle v,u\rangle=0$ also $\left|\langle v,P_t^* u\rangle\right|^2$ vanishes at $t=0$ and
\[
\left\langle v,\mathcal{T}_t(|u\rangle\langle u|)v\right\rangle
\geq \sum_{j,k=1}^{d^2-1}c_{jk}\int_{0}^t
\left\langle v,\mathcal{T}_s(|F_j^*P_{t-s}^*u\rangle\langle F_k^*P_{t-s}^*u|)v\right\rangle \hbox{\rm d}s
\]
for all $t\geq 0$. Both sides of the above inequality vanish at $t=0$ and so
\[
\frac{\hbox{\rm d}}{\hbox{\rm d}t} \left\langle v,\mathcal{T}_t(|u\rangle\langle u|)v\right\rangle\Big|_{t=0}
\kern-2truept  \geq \kern-4truept
\sum_{j,k=1}^{d^2-1} \kern-2truept c_{kj} \left\langle v,|F_j^* u\rangle\langle F_k^* u|v\right\rangle
= \kern-4truept \sum_{j,k=1}^{d^2-1} \kern-2truept c_{kj}
\left\langle F_j v, u\right\rangle \left\langle u, F_k v\right\rangle.
\]
The right-hand side is strictly positive because the matrix $(c_{jk})_{1\leq j,k\leq d}$ is strictly positive
definite and $\left\langle F_k v, u\right\rangle\not=0$ for some $k$ because matrices $F_k$ (with the identity
matrix) form a basis of $M_d(\mathbb{C})$. It follows that, also in the case where $\langle v,u\rangle=0$,
we can find $t_{u,v}>0$ such that $\left\langle v,\mathcal{T}_t(|u\rangle\langle u|)v\right\rangle>0$ for
all $t\in ]0,t_{u,v}]$.

For all  $(u,v)$ in $S_d\times S_d$ (the product of two copies of the unit sphere of $\mathbb{C}^d$),
by continuity, we can find an open neighbourhood $\mathcal{U}(u,v)$ of $(u,v)$ such that
$\left\langle v',\mathcal{T}_t(|u'\rangle\langle u'|)v'\right\rangle>0$ for all
$t\in]0,t_{u,v}/2]$ and $(v',u')\in \mathcal{U}(u,v)$. The family of open sets $\mathcal{U}(u,v)$
is a covering of the compact set $S_d\times S_d$, therefore we can extract a finite subcovering
$(\mathcal{U}(u_i,v_i))_{1\leq i \leq n}$. Considering $t_0=\min_{1\leq i \leq n}t_{u_i,v_i}/2$
we have
\[
\left\langle v,\mathcal{T}_t(|u\rangle\langle u|)v\right\rangle>0
\]
for all unit vectors $u,v\in\mathbb{C}^d$ and $t\in]0,t_0]$ so that $\mathcal{T}_t$ is positivity improving.
One can find, in particular, an $\eta>0$ such that $\mathcal{T}_{t_0}(|u\rangle\langle u|)>\eta\unit$
for all unit vector $u\in\mathbb{C}^d$ therefore, for all $t>t_0$, we have
\[
\mathcal{T}_{t}(|u\rangle\langle u|)= \mathcal{T}_{t-t_0}\left(\mathcal{T}_{t_0}(|u\rangle\langle u|)\right)
>\eta \mathcal{T}_{t-t_0}(\unit) = \eta \unit.
\]
This completes the proof. \hfill $\square$

\section{The Kossakowski matrix of a Gaussian QMS}\label{sect:Koss-mat}

Let $\mathsf{h}$ be the Boson Fock space $\Gamma(\mathbb{C}^d)$, isometrically isomorphic to
$\Gamma(\mathbb{C})\otimes\cdots\otimes\Gamma(\mathbb{C})$, and fix the canonical orthonormal basis
$(e(n_1,\ldots,n_d))_{n_1,\ldots,n_d\geq 0}$, with $e(n_1,\ldots,n_d)=e_{n_1}\otimes\ldots\otimes e_{n_d}$.
Let $a_j, a_j^{\dagger}$ be the creation and annihilation operator of the Fock representation of the $d$-dimensional CCR
\begin{eqnarray*}
a_j\,e(n_1,\ldots,n_d)& =& \sqrt{n_j}\,e(n_1,\ldots,n_{j-1},n_j-1,\ldots,n_d),\\
a_j^{\dagger}\,e(n_1,\ldots,n_d)&=&\sqrt{n_j+1}\,e(n_1,\ldots,n_{j-1},n_j+1,\ldots,n_d),
\end{eqnarray*}
satisfying the CCR $[a_j,a_k^{\dagger}]=\delta_{jk}\unit $.
 Define the coherent (also called exponential) vector $e(g)$ associated with $g$ by
\[
 e_g=\sum_{n\in \mathbb{N}^d}\frac{g_1^{n_1}\cdots g_d^{n_d}}{\sqrt{n_1!\cdots n_d!}}\,e(n_1,\ldots,n_d)
\]
and define creation and annihilation operators
 \[a(v)e(g)=\langle v,g \rangle e(g), \quad a^{\dagger}(v)e(g)=\frac{d}{d\varepsilon}e(g+\varepsilon u)|_ {\varepsilon=0}\]
for all $u\in\mathbb{C}^d$ and note that
\[
 a(v)=\sum\limits_{j=1}^d\overline{v}_ja_j, \quad a^{\dagger}(u)=\sum\limits_{j=1}^d u_ja_j^{\dagger}
\]
for all $u^{\hbox{\rmseven T}}=(u_1,\ldots,u_d)$, $v^{\hbox{\rmseven T}}=(v_1,\ldots,v_d)\in\mathbb{C}^d$.

The above operators are obviously defined on the linear manifold $D$ spanned by the elements
$(e(n_1,\ldots,n_d))_{n_1,\ldots,n_d\geq0}$ of the canonical orthonormal basis of $\mathsf{h}$.

Gaussian quantum Markov semigroups (QMS) have an unbounded generator (\cite{Po2021} and the references therein) and
GKLS representation is not defined on all bounded operators $x$ as (\ref{eq:GKLSbd}).
To keep in mind this point, we refer to the following $\mathcal{L}$ as  pre-generator in a generalized
Gorini–Kossakowski–Lindblad-Sudarshan (GKLS) form
\begin{equation}\label{eq:GKLS}
\mathcal{L}(x) = \mi\left[ H, x\right]
-\frac{1}{2}\sum_{\ell=1}^m \left( L_\ell ^*L_\ell\, x - 2 L_\ell^* x L_\ell + x\, L_\ell ^*L_\ell\right).
\end{equation}
Here $1 \leq m \leq 2d$,
\begin{eqnarray}
		H \kern-4truept &=& \kern-6truept
          \sum_{j,k=1}^d \kern-2truept \left( \Omega_{jk} a_j^\dagger a_k + \frac{\kappa_{jk}}{2} a_j^\dagger a_k^\dagger
        + \frac{\overline{\kappa_{jk}}}{2} a_ja_k \right) \kern-2truept
        + \kern-2truept \sum_{j=1}^d \kern-2truept\left( \frac{\zeta_j}{2}a_j^\dagger + \frac{\bar{\zeta_j}}{2} a_j
        \kern-2truept\right)\kern-2truept, \label{eq:H}\\
		L_\ell \kern-4truept &=& \kern-6truept
       \sum_{k=1}^d \left( \overline{v_{\ell k}} a_k + u_{\ell k}a_k^\dagger\right)
                 = a( v_{\ell \bullet}) + a^\dagger ( u_{\ell \bullet}), \label{eq:Lell}
\end{eqnarray}
with $\Omega:=(\Omega_{jk})_{jk} = \Omega^*$, $\kappa:= (\kappa_{jk})_{jk}= \kappa^{\hbox{\texttt{\scriptsize T}}} \in M_d(\mathbb{C})$,
$V=(v_{\ell k})_{\ell k}, U=(u_{\ell k})_{\ell k} \in M_{m\times d}(\mathbb{C})$, $\zeta=(\zeta_j)_j \in \mathbb{C}^d$,
$v_{\ell \bullet}$ and $u_{\ell \bullet}$ denote the $\ell$-th row of matrices $V$ and $U$.\\
We assume that either $U$ or $V$ is non-zero so that the Kossakowski matrix we will find is non-zero.
Moreover we also choose the number of Kraus' operators $L_\ell$ (namely the parameter $m$) according to the following definition.
	
\begin{definition}
A GKLS representation is \emph{mimimal} if the number $m$ in \emph{(\ref{eq:GKLS})} is minimum.
\end{definition}

By the linear independence of  $(a_j,a^\dagger_j)_{1\leq j \leq d}$  we have the following

\begin{proposition}
The Gaussian pre-generator \emph{(\ref{eq:GKLS})} has a minimal GKLS representation if and only if
\begin{equation} \label{eq:minimalCond}
\hbox{\rm ker}\left(V^{*}\right)\cap \hbox{\rm ker}\left(U^{\hbox{\rmseven T}}\right)=\{0\}.
\end{equation}
 \end{proposition}
We refer to \cite{AFP-MJM} Prop. 2 for the proof.
Condition (\ref{eq:minimalCond}) will be assumed throughout the paper.

It is known that, interpreting (\ref{eq:GKLS}) as a quadratic form
$\pounds(x)$ with domain $D\times D$ for all bounded operator $x$, one can construct a QMS with
unbounded generator $\mathcal{L}$ (see e.g. \cite{AFP-MJM} Appendix A).
More precisely, let $G, G_0$ be the closure of operators  defined on $D$ by
\[
G = -\mathrm{i} H -\frac{1}{2}\sum_{\ell=1}^{2d} L_\ell^*L_\ell, \qquad
G_0 = -\frac{1}{2}\sum_{\ell=1}^{2d} L_\ell^*L_\ell.
\]

\begin{proposition} \label{prop:Ggenerates}
The operator $G$ is the infinitesimal generator of a strongly continuous contraction semigroup on $\mathsf{h}$
and $D$ is a core for this operator. The operator $G_0$ is negative self-adjoint.
\end{proposition}

For all $x\in\mathcal{B}(\mathsf{h})$ consider the quadratic form with domain $D\times D$
\begin{eqnarray}\label{eq:Lform}
\texttt{\textrm{\pounds}}(x)  [v,u] &= & \mi\left\langle{H v},{x u}\right\rangle - \mi\left\langle{v},{x H u}\right\rangle \\
			&- & \frac{1}{2} \sum_{\ell=1}^{2d} \left( \left\langle{v},{xL_\ell^*L_\ell u}\right\rangle
              -2\left\langle{L_\ell v},{xL_\ell u}\right\rangle
         + \left\langle{L_\ell^* L_\ell v},{x u}\right\rangle \right) \nonumber
\end{eqnarray}

We can prove the following (see \cite{AFP-MJM} Appendix A)
\begin{theorem}\label{th:G-QMS!}
There exists a unique QMS, $\mathcal{T}=(\mathcal{T}_t)_{t\geq 0}$ such that, for all $x\in\mathcal{B}(\mathsf{h})$ and
$v,u\in D$, the function $t\mapsto  \left\langle{v},{\mathcal{T}_t (x) u}\right\rangle  $ is differentiable and
\[
	\left\langle{v},{\mathcal{T}_t (x) u}\right\rangle
= \left\langle{v},{x u}\right\rangle
 + \int_0^t \texttt{\textrm{\pounds}}(\mathcal{T}_s(x)) [v,u] \hbox{\rm d}s
 \qquad \forall\, t\geq 0.
\]
The domain of the generator consists of $x\in\mathcal{B}(\mathsf{h})$ for which the quadratic form
$\texttt{\textrm{\pounds}}(x)$ is represented by a bounded operator.
\end{theorem}

Let $\mathcal{L}_0(x)=\mathcal{L}(x)-\mi[H,x]$ be the pre-generator $\mathcal{L}$ (\ref{eq:GKLS}) without the
Hamiltonian term.  A straightforward computation yields
\begin{eqnarray*}
  \mathcal{L}_0(x) \kern-4truept & \kern-2truept = \kern-2truept & \kern-4truept
  \frac{1}{2}\sum_{\ell=1}^m \left([L_\ell^*,x]L_\ell + L_\ell^*[x,L_\ell] \right) \\
    \kern-4truept & \kern-2truept = \kern-2truept & \kern-6truept
   \frac{1}{2} \kern-4truept
   \sum_{j,k=1}^d \kern-4truept \Big( \kern-2truept (V^{\hbox{\scriptsize T}} \overline{V})_{jk}\left([a^\dagger_j,x]a_k
    \kern-3truept + \kern-2truept a^\dagger_j[x,a_k] \kern-1truept \right)
   \kern-2truept + \kern-2truept (V^{\hbox{\scriptsize T}} U)_{jk}\kern-2truept\left([a^\dagger_j,x]a^\dagger_k \kern-1truept
   + \kern-1truept a^\dagger_j [x,a^\dagger_k]\right)\\
   \kern-4truept & \kern-2truept   \kern-2truept & \kern-4truept  + (U^* \overline{V})_{jk}\left([a_j,x]a_k + a_j[x,a_k]\right)
   + (U^*U)_{jk}\left([a_j,x]a^\dagger_k+a_j[x,a^\dagger_k]\right)\Big)\\
\end{eqnarray*}
As a result, $\mathcal{L}_0$ (with respect to the
``basis'' $a_1,\dots,a_d,a^\dagger_1,\dots, a^\dagger_d$) is determined by a $2d\times 2d$ matrix.

\begin{definition}\label{def:Koss-mat}
We call Kossakowski matrix of the QMS generated by \emph{(\ref{eq:GKLS})} the $2d\times 2d$ matrix
with four $d\times d$ blocks
\[
 \mathbb{K}= \left[ \begin{array}{cc}
 V^{\hbox{\emph{\scriptsize T}}} \overline{V} & V^{\hbox{\emph{\scriptsize T}}} U \\    U^* \overline{V} & U^*U
 \end{array}\right]
\]
\end{definition}

\noindent{\bf Remark.} It is well-known (see e.g. \cite{Partha} Theorem 30.16 p.271) that the operators
$L_\ell$ in a GKLS representation are not unique. Indeed, disregarding addition of multiples of the
identity operator, one can choose other operators $L'_l = \sum_{j=1}^m r_{lj}L_j$ for a unitary
matrix $(r_{lj})_{1\leq l,j\leq m}$ and get the same $\mathcal{L}_0$. A straightforward computation
shows that $\mathbb{K}$ is invariant for such a transformation.

Throughout the paper we assume $\mathbb{K}>0$. As a consequence, since $\mathbb{K}$ can be decomposed as
the product of a $2d\times m$ matrix and its adjoint
\[
\mathbb{K}=
\left[\begin{array}{c} V^{\hbox{\scriptsize T}} \\ U^*\end{array} \right] \left[\begin{array}{cc} \overline{V} & U\end{array} \right] =
\left[\begin{array}{cc} V^{\hbox{\scriptsize T}}\overline{V} & V^{\hbox{\scriptsize T}}U \\ U^*\overline{V} & U^*U \end{array} \right] >0
\]
with $m\leq 2d$, strict positivity of $\mathbb{K}$ implies $m=2d$.

\noindent{\bf Remark.} The matrix $\mathbb{K}$ is not invariant for  Bogoliubov transformations but
it is not difficult to show that condition $\mathbb{K}>0$ is. Indeed, a general Bogoliubov transformation is
invertible and, denoting by $\mathfrak{a},\mathfrak{a}^\dagger$ vectors $[a_1,\dots a_d]^{\hbox{\scriptsize T}}$, 
$[a_1^\dagger,\dots a_d^\dagger]^{\hbox{\scriptsize T}}$ and $\mathfrak{b},\mathfrak{b}^\dagger$ their 
Bogoliubov transformed
\begin{eqnarray*}
\left[\begin{array}{c} \mathfrak{a} \\ \mathfrak{a}^\dagger\end{array} \right]
& = & \left[\begin{array}{cc} E^{\hbox{\scriptsize T}} & F^{\hbox{\scriptsize T}} \\ F^* & E^* \end{array} \right]
\left[\begin{array}{c} \mathfrak{b} \\ \mathfrak{b}^\dagger\end{array} \right]
\end{eqnarray*}
for two $d\times d$ complex matrices satisfying
\begin{equation}\label{eq:Bgl-cnd}
E^*E-F^*F=\unit, \qquad E^TF - F^T E =0.
\end{equation}
and so $\mathbb{K}$ becomes
\[
\left[\begin{array}{cc} \overline{E} & F\\ \overline{F} & E \end{array} \right]\cdot
\left[\begin{array}{cc} V^{\hbox{\scriptsize T}}\overline{V} & V^{\hbox{\scriptsize T}}U \\ U^*\overline{V} & U^*U \end{array} \right]
\cdot \left[\begin{array}{cc} E^{\hbox{\scriptsize T}} & F^{\hbox{\scriptsize T}} \\ F^* & E^* \end{array} \right].
\]

\medskip
In this section we prove that a Gaussian QMS with a strictly positive Kossakowski matrix is irreducible.
We begin by some preliminary results on operator domains that are needed in the sequel.

\begin{lemma}\label{lem:DomG=DomN}
Suppose $\mathbb{K}>0$ and let $\epsilon_0>0$ be the smallest eigenvalue of $\mathbb{K}$.
For all $\xi\in\hbox{\rm Dom}(N)$ we have
\begin{equation}\label{eq:G01st}
\left\langle\xi, -2G_0\xi\right\rangle \geq \epsilon_0 \left\langle \xi, (2N+d \emph{\unit} )\xi\right\rangle.
\end{equation}
\end{lemma}

\noindent{\it Proof.}
Note that, for all $\xi\in D$, denoting by $a^\sharp \xi$ the vector
$[a_1\xi,\dots,a_d\xi,$ $ a_1^\dagger\xi,\dots, a_d^\dagger\xi]^{\hbox{\scriptsize T}}$
in the direct sum $\mathsf{h}_{2d}$ of $2d$ copies of $\mathsf{h}$ and by
$\langle\cdot,\cdot\rangle_{2d}$ the corresponding natural scalar product, we have
\[
\left\langle\xi, -2G_0\xi\right\rangle
=\sum_{\ell=1}^{2d}\left\langle  L_\ell\xi, L_\ell\xi\right\rangle
= \left\langle a^\sharp \xi, \mathbb{K}\, a^\sharp \xi\right\rangle_ {2d}.
\]
The following inequality
\[
\left\langle\xi, -2G_0\xi\right\rangle \geq  \epsilon_0 \left\langle a^\sharp \xi,  a^\sharp \xi\right\rangle_ {2d}
= \epsilon_0 \left\langle \xi, (2N+d \unit )\xi\right\rangle.
\]
is immediate and (\ref{eq:G01st}) is proved.
\hfill $\square$

\begin{theorem}\label{th:DomG=DomN}
If $\mathbb{K}>0$ then there exist constants $c,c_0>0$ such that
\begin{eqnarray}
  \epsilon_0^2\left\Vert N\xi\right\Vert^2 & \leq & 2\left\Vert G_0\xi\right\Vert^2 + c_0\Vert \xi\Vert^2 \label{eq:NrbG0} \\
  \epsilon_0^2 \left\Vert N\xi\right\Vert^2 &  \leq & 2\left\Vert G  \xi\right\Vert^2 + c\kern1truept\Vert \xi\Vert^2 \label{eq:NrbG}
\end{eqnarray}
for all $\xi\in D$. In particular $\hbox{\rm Dom}(G)=\hbox{\rm Dom}(N)=\hbox{\rm Dom}(G_0)$.
\end{theorem}

\noindent{\it Proof.}
Clearly $\hbox{\rm Dom}(N)\subseteq \hbox{\rm Dom}(G)$ and
$\hbox{\rm Dom}(N)\subseteq \hbox{\rm Dom}(G_0)$. First we prove the opposite inclusion for
$\hbox{\rm Dom}(G_0)$.  For all $\xi\in D$, we have also
\begin{eqnarray*}
\left\Vert 2G_0\xi\right\Vert^2
& = & \sum_{j,\ell=1}^{2d}\left\langle  L_j^*L_j\xi, L_\ell^*L_\ell\xi\right\rangle \\
& = & \sum_{j,\ell=1}^{2d}\left\langle  L_\ell\xi, L_j^*L_j L_\ell\xi\right\rangle
+ \sum_{j,\ell=1}^{2d}\left\langle  [L_\ell,L_j^*L_j]\xi, L_\ell\xi\right\rangle.
\end{eqnarray*}
A straightforward computation shows that commutators $[L_\ell,L_j^*L_j]$ are linear
in $a_h,a^\dagger_k$, therefore one can find a constant $k_1>0$ such that
$\left| \left\langle [L_\ell,L_j^*L_j]\xi, L_\ell\xi\right\rangle\right| \leq k_1 \Vert (N+ \unit )^{1/2}\xi \Vert^2$
for all $j,\ell$ and we find the inequality
\begin{eqnarray*}
\left\Vert 2G_0\xi\right\Vert^2
& \geq & \sum_{j,\ell=1}^{2d}\left\langle  L_\ell\xi, L_j^*L_j L_\ell\xi\right\rangle
- (2d)^2 k_1 \left\langle \xi, (N+ \unit )\xi\right\rangle \\
& = & \sum_{\ell=1}^{2d}\left\langle  L_\ell\xi, (-2G_0) L_\ell\xi\right\rangle
- (2d)^2 k_1 \left\langle \xi, (N+ \unit )\xi\right\rangle.
\end{eqnarray*}
By the Young inequality $rs\leq \epsilon_0^2 r^2 + s^2/(4\epsilon_0^2)$ we have
\[
(2d)^2 k_1 \left\langle \xi, (N+ \unit )\xi\right\rangle
\leq (2d)^2 k_1\Vert\xi\Vert\cdot \Vert (N+\!\!\unit )\xi\Vert
\leq \epsilon_0^2 \Vert (N+ \unit )\xi\Vert ^2 + \frac{d^4 k_1^2}{\epsilon_0^2}\Vert\xi\Vert^2
\]
and so, by  (\ref{eq:G01st}) and the Schwarz inequality, we find
\begin{equation}\label{eq:G0geq}
\left\Vert 2G_0\xi\right\Vert^2
\geq   \epsilon_0\sum_{\ell=1}^{2d}\left\langle  L_\ell\xi, (2N+d\!\unit ) L_\ell\xi\right\rangle
-  \epsilon_0^2  \Vert (N+ \unit )\xi\Vert ^2 - \frac{d^4 k_1^2}{\epsilon_0^2}\Vert\xi\Vert^2.
\end{equation}
Write $N= \sum_{j=1}^d a^\dagger_j a_j$. Commuting $a_j$  with
$L_\ell$ and noting that commutators $[a_j,L_\ell]$ are multiples of the identity operator
one can find another constant $k_2>0$ such that the first term in the right-hand side,
in turn, is bigger or equal than
\begin{eqnarray*}
& & 2\epsilon_0\sum_{j=1}^d \sum_{\ell=1}^{2d}\left\langle  L_\ell a_j \xi, L_\ell a_j\xi\right\rangle
+ d\epsilon_0\sum_{\ell=1}^{2d}\left\langle  L_\ell \xi, L_\ell \xi\right\rangle
- k_2 \Vert (N+ \unit )^{1/2}\xi\Vert ^2 \\
& = & 2\epsilon_0\sum_{j=1}^d
\left(\left\langle   a_j \xi, (-2G_0)a_j\xi\right\rangle
+ d\epsilon_0\left\langle  \xi, (-2G_0) \xi\right\rangle\right)
- k_2 \Vert (N+ \unit )^{1/2}\xi\Vert ^2.
\end{eqnarray*}
Another application of (\ref{eq:G01st}) and a computation with the CCR yield
\begin{eqnarray*}
\epsilon_0\sum_{\ell=1}^{2d}\left\langle  L_\ell\xi, (2N+d\!\unit ) L_\ell\xi\right\rangle
& \geq & 2\epsilon_0^2\sum_{j=1}^d\left\langle a_j\xi, (2N+d\!\unit ) a_j \xi\right\rangle \\
&  & \kern-3truecm + \ d\epsilon_0^2 \left\langle   \xi, (2N+d\!\unit ) \xi\right\rangle
- k_2 \Vert (N+ \unit )^{1/2}\xi\Vert ^2 \\
& & \kern-3truecm =  \epsilon_0^2 \left\langle   \xi, (4N^2+2(d-1)N+d^2 \unit ) \xi\right\rangle
- k_2 \Vert (N+ \unit )^{1/2}\xi\Vert ^2
\end{eqnarray*}
Plugging this inequality into (\ref{eq:G0geq}) and noting  that
$\Vert (N+ \unit )^{1/2}\xi\Vert ^2$ is not bigger than $ \Vert N^{1/2}\xi\Vert ^2 + 2 \Vert \xi\Vert ^2 $ we find
\[
   \left\Vert 2G_0\xi\right\Vert^2
   \geq  3\epsilon_0^2 \Vert N\xi\Vert^2
   +\left((2d \kern-1truept -\kern-1truept 3)\epsilon^2_0 \kern-1truept - \kern-1truept k_2\right) \kern-1truept \Vert N^{1/2}\xi\Vert^2
   + \left(d^2\epsilon^2_0-d^4k_1^4/\epsilon^2_0\kern-3truept - \kern-2truept 2k_2\right)\kern-2truept\Vert \xi\Vert^2
\]
and, finally, in the case where $(2d-3)\epsilon^2_0-k_2$ is strictly negative, by another application of the
Young inequality $rs\leq \epsilon_0^2 r^2 + s^2/(4|(2d-3)\epsilon^2_0-k_2|)$ we can find another
constant $k_3>0$ such that
\begin{equation}\label{eq:NrbwrG0}
2\epsilon_0^2\left\Vert N\xi\right\Vert^2 \leq \left\Vert 2 G_0\xi\right\Vert^2 + k_3 \Vert \xi\Vert^2.
\end{equation}
Th inequality (\ref{eq:NrbG0}) immediately follows and allows us to show that $\hbox{\rm Dom}(G_0)\subseteq\hbox{\rm Dom}(N)$.
Indeed, if $\xi\in \hbox{\rm Dom}(G_0)$, then there exists a sequence $(\xi_n)_{n\geq 1}$ in
$D$ converging in norm to $\xi$ such that also $(G_0\xi_n)_{n\geq 1}$ converges in norm to $\xi$.
It follows that also $(N\xi_n)_{n\geq 1}$ converges and so $\xi$ belongs to $\hbox{\rm Dom}(N)$.

We can prove that $\hbox{\rm Dom}(G)\subseteq\hbox{\rm Dom}(N)$ in a similar way.
First note that, for all $\xi\in D$,
\begin{eqnarray*}
\left\Vert G\xi\right\Vert^2
& = & \left\langle \xi, \left(G_0^2+H^2\right) \xi\right\rangle
+\left\langle \xi, \mathrm{i}[H,G_0] \xi\right\rangle.
\end{eqnarray*}
A straightforward computation shows that the commutator $[H,G_0]$ is a second order polynomial
in $a_j,a^\dagger_k$, therefore one can find a constant $k_4>0$ such that
$\left| \left\langle \xi, \mathrm{i}[H,G_0] \xi\right\rangle\right| \leq k_4 \Vert (N+\unit )^{1/2} \xi\Vert^2$
and we find the inequality
\begin{eqnarray}\label{eq:GqbtN}
\left\Vert G\xi\right\Vert^2 & \geq &
\left\langle \xi, \left(G_0^2+H^2\right) \xi\right\rangle
- k_4 \left\langle \xi, (N+\unit ) \xi\right\rangle \nonumber \\
 & \geq &
\left\langle \xi, G_0^2 \xi\right\rangle
- k_4 \left\langle \xi, (N+\unit ) \xi\right\rangle.
\end{eqnarray}
Applying (\ref{eq:NrbwrG0}) and the Young inequality with suitable weights we get (\ref{eq:NrbG}).
The proof of the inclusion $\hbox{\rm Dom}(G)\subseteq\hbox{\rm Dom}(N)$ now follows
the same line of the proof of the previous inclusion $\hbox{\rm Dom}(G_0)$ $\subseteq\hbox{\rm Dom}(N)$.
\hfill $\square$

\smallskip
The following result (see Theorem III.1 \cite{FFRR02}) characterizes subharmonic projections for a QMS.
\begin{theorem}\label{th:subpr-invsp}
A projection $p$ is subharmonic for $\mathcal{T}$ if and only if the range $\hbox{\rm Rg}(p)$ of $p$ is invariant
for the operators $P_t$ ($t\ge 0$) of the strongly continuous contraction semigroup on $\mathsf{h}$ generated
by $G$ and $L_\ell u = p L_\ell u$, for all $u\in \hbox{\rm Dom}(G)\cap \hbox{\rm Rg}(p)$, and all $\ell \ge 1$.
\end{theorem}

It is worth noticing that $\hbox{\rm Dom}(G)\cap \hbox{\rm Rg}(p)$ is dense in
$\hbox{\rm Rg}(p)$ by a well-known property of subspaces invariant under maps $P_t$ of a strongly
continuous semigroup.

\begin{theorem}\label{th:irreduc}
If $\mathbb{K}>0$ then the QMS with generalized GKSL generator associated with operators $H,L_\ell$
as in  (\ref{eq:H}), (\ref{eq:Lell}) is irreducible.
\end{theorem}
\noindent{\it Proof.}
Let $\mathcal{V}$ be a nonzero closed subspace of $\mathsf{h}$ which is invariant for
the contraction operators $P_t$ of the semigroup generated by $G$ and
$L_\ell\left(\hbox{\rm Dom}(G)\cap \mathcal{V}\right)\subseteq  \mathcal{V}$ for $\ell=1,2$.

By the linear independence of $L_1,\dots,L_{2d}\,$, since $\hbox{\rm Dom}(G)=\hbox{\rm Dom}(N)$, we have also
\begin{eqnarray*}
a_j\left(\hbox{\rm Dom}(N)\cap\mathcal{V}\right) \subseteq \hbox{\rm Dom}(N^{1/2})\cap\mathcal{V}
&  & \kern-8truept a^\dagger_k\left(\hbox{\rm Dom}(N)\cap\mathcal{V}\right) \subseteq \hbox{\rm Dom}(N^{1/2})\cap\mathcal{V} \\
a^\dagger_k a_j\left(\hbox{\rm Dom}(N)\cap\mathcal{V}\right) \subseteq \mathcal{V}
&  & \kern-8truept
a_j a^\dagger_k \left(\hbox{\rm Dom}(N)\cap\mathcal{V}\right) \subseteq \mathcal{V}
\end{eqnarray*}
hence, denoting by $p$ the orthogonal projection onto $\mathcal{V}$,
\[
p^\perp a_j p =0 = p\, a_j p^\perp  \qquad p^\perp a^\dagger_k  p =0 = p\, a^\dagger_k p^\perp
\]
on $\hbox{\rm Dom}(N)\cap\mathcal{V}$ for all $j,k$ and, left multiplying by $a^\dagger$ the first identity,
\[
p^\perp a^\dagger_j a_j p =0 = p\, a^\dagger_j a_j p^\perp.
\]
It follows that, for all $t\geq 0$, $n>0$,
$\left( \unit  + tN/n\right)$ commutes with $p$ and,
left and right multiplication by the resolvent $\left( \unit + tN/n\right)^{-1}$ yields
\[
p \left(\unit + tN/n\right)^{-1}= \left( \unit + tN/n\right)^{-1}p.
\]
Multiplying both sides by $\left( \unit + tN/n\right)^{-n+1}$ we have
\[
p\left(\!\! \unit + tN/n\right)^{-n}
=\left( \unit + tN/n\right)^{-n}p
\]
for all $t\geq 0$. Taking the limit as $n$ tends to infinity, by the Hille-Yosida theorem
(\cite{BrRo} Theorem 3.1.10 p.371) we get the commutation
\begin{equation}\label{eq:p-etN-p=etN-p}
p\,\hbox{\rm e}^{-tN}=\hbox{\rm e}^{-tN}p \qquad \forall t\geq 0.
\end{equation}

Let $v\in\mathcal{V}$, $v\not=0$ with expansion in the canonical basis
\[
v = \sum_{|n|\ge n_0} v_n e_n
\]
where $n_0$ is the minimum of $|n|=n_1+\dots+n_d$ over multindexes $n=(n_1,\dots, n_d)$
for which $v_n\not=0$. Clearly, by (\ref{eq:p-etN-p=etN-p}), $\hbox{\rm e}^{-t N} v \in\mathcal{V}$
for all $t\ge 0$ and so
\[
\hbox{\rm e}^{n_0 t }\hbox{\rm e}^{-t N} v =
\sum_{|n|\ge n_0} \hbox{\rm e}^{-(|n|-n_0)t} v_n e_n
= \sum_{|n| = n_0}v_{n} e_{n}+\sum_{|n|>n_0} \hbox{\rm e}^{-(|n|-n_0)t} v_n e_n\in\mathcal{V}
\]
for all $t\ge 0$. Taking the limit at $t\to +\infty$, we find $\sum_{|n|= n_0} v_n e_n\in\mathcal{V}$.
Acting this non-zero vector with operators $a_j$ and $a^\dagger_k$  we can immediately show that
the vacuum vector $e_0$ belongs to $\mathcal{V}$. It follows that each
\[
e_n = (n_1!\cdots n_d\,!)^{-1/2} a^{\dagger\,{n_1}}_1\cdots a^{\dagger\,{n_d}}_d e_0
\]
belongs to $\mathcal{V}$ and the proof is complete. \hfill $\square$


\section{Positivity Improving Gaussian QMSs}\label{sect:pos-impr}

Strict positivity of the Kossakowski matrix, along with additional conditions on $H$, imply that semigroup $(P_t)_{t\geq 0}$
generated by $G$ is analytic. This is a useful step in the analysis of a QMS with unbounded generator because operators
$P_t$ ($t>0$) are well-behaved with respect to operator domains (e.g., for all $t>0$,  $P_t$ maps the whole Hilbert space
in $\hbox{\rm Dom}(N^n)$ for all $n>0$) and one can make sense of several operator compositions as in the case of
bounded operators.


In this section we show that strict positivity of $\mathbb{K}$ implies that a Gaussian QMS is
positivity improving if we assume that the semigroup $(P_t)_{t\geq 0}$ is analytic.
We begin by a technical lemma

\begin{lemma}\label{lem:DGn=DNn}
If $\mathbb{K}>0$ then  $\hbox{\rm Dom}(G^n)= \hbox{\rm Dom}(N^n)$ or all $n\geq 0$.
\end{lemma}

\noindent{\it Proof.} Clearly $\hbox{\rm Dom}(N^n) \subseteq \hbox{\rm Dom}(G^n)$.
Arguing as in the proof of Theorem \ref{th:DomG=DomN} we will prove by induction that
there exists constants $\alpha,\beta_{kh}>0$ ($0\leq h\leq k$)  such that
\begin{equation}\label{eq:Nn-rb-Gn}
\left\Vert (N+ \unit)^n \xi\right\Vert \leq \alpha^n \left\Vert G^n \xi\right\Vert +
\sum_{k=0}^{n-1}\beta_{nk} \left\Vert G^k\xi\right\Vert
\end{equation}
for all $n\geq 0$ and all $\xi\in D$.

For $n=1$ it immediately follows from the inequality (\ref{eq:NrbG}).
Suppose that (\ref{eq:Nn-rb-Gn}) has been established for an integer $n\geq 1$.
Put $X=N+ \unit$ to simplify the notation. Writing
$X^{n+1}\xi= X^n X\xi$ we have
\begin{equation}\label{eq:Nn1Gn1}
\left\Vert X^{n+1} \xi\right\Vert
\leq  \alpha^n\left\Vert G^n X \xi\right\Vert + \sum_{k=0}^{n-1}\beta_{nk}  \left\Vert  G^k X \xi \right\Vert
\end{equation}
Writing $G^k X= X G^k + [G^k,N]$, since the commutator $[N, G^k]$ is a polynomial of order $2k$ in
$a_j,a^\dagger_h$, and so we can find a constant $\gamma_k$ such that $\left\Vert [N, G^k] \xi\right\Vert \leq \gamma_k\left\Vert X^k\xi\right\Vert$,
for all $k=1,\dots, n$, we have the inequality
\begin{eqnarray*}
\left\Vert G^k X \xi\right\Vert
& \leq &  \left\Vert X G^k \xi\right\Vert +  \left\Vert [G^k, N] \xi\right\Vert \\
& \leq &   \alpha\left\Vert G^{k+1}\xi\right\Vert + \beta_{01}\Vert G^k \xi \Vert
+ \gamma_k  \left\Vert  X^k  \xi \right\Vert \\
& \leq & \alpha \left\Vert G^{k+1}\xi\right\Vert
+ (\alpha\gamma_k +\beta_{01}) \left\Vert G^k \xi\right\Vert + \alpha \gamma_k
\sum_{h=0}^{k-1} \beta_{kh} \left\Vert G^h\xi\right\Vert .
\end{eqnarray*}
Plugging this inequality in (\ref{eq:Nn1Gn1}) a straightforward computation yields (\ref{eq:Nn-rb-Gn})
for $n+1$.

Equation (\ref{eq:Nn-rb-Gn}) allows us to check that $\hbox{\rm Dom}(G^n) \subseteq \hbox{\rm Dom}(N^n)$ for
all $n\geq 0$ as in the proof of Theorem \ref{th:DomG=DomN}. \hfill $\square$

\smallskip
Lemma \ref{lem:DGn=DNn} implies that $L_\ell (\hbox{\rm Dom}(G^k))\subseteq \hbox{\rm Dom}(G^{k-1})$ for
all $k\geq 1$ and we can apply Theorem 2 in \cite{GlHaNa}. We recall this result.
Below $\delta_{G}^{m}(L_{\ell})$ denote the $m$-times iterated commutator of $G$ and $L_\ell$.
More precisely, one can note that the iterated commutators $[G,[\dots,[G,L]]]$, defined on
the domain $D$, are linear combinations of $a_j,a^\dagger_k$ and therefore can be extended to
closed operators defined on $\hbox{\rm Dom}(N^{1/2})$ denoted by $\delta_{G}^{m}(L_{\ell})$.

\begin{theorem}\label{th:GlHaNa}
Let $(\mathcal{T}_t)_{t \geq 0}$ the QMS on $\mathcal{B}(\mathsf{h})$ associated with operators $H,L_\ell$
as in (\ref{eq:H}), (\ref{eq:Lell}). Suppose that $G$ generates an analytic semigroup in
a sector $\left\{\, z \in \mathbb{C}-\{0\} \, \mid\,  |\hbox{\rm Arg }(z)| < \theta \,\right\}$
for some $\theta \in ]0,\pi/2[$ and, moreover, that
\[
L_{\ell}(\Dom(G^{k})) \subseteq \Dom(G^{k-1})
\]
for all $k> 0$. For all state $\omega=\sum_{j \in J}\omega_{j}|\psi_{j}\rangle \langle \psi_{j}|,$
with $\omega_{j}> 0$ for all $j\in J$ and all $t \geq 0,$ the support
$\mathcal{S}_{t}(\omega)$ of the state $\mathcal{T}_{*t}(\omega)$
is the closure of linear manifold generated by vectors
\begin{equation}\label{eq:suport3}
P_{t}\psi_j, \
 \delta_{G}^{m_{1}}(L_{\ell_1})\delta_{G}^{m_{2}}(L_{\ell_2})\cdots
 \delta_{G}^{m_{n}}(L_{\ell_n})P_{t}\psi_{j}
\end{equation}
for all $j\in J$, $n \geq 1,$ $ m_{1},\cdots,m_{n} \geq 0$ and $\ell_{1},\cdots, \ell_{n} \geq 1$.
\end{theorem}

\begin{theorem}\label{th:pos-imp}
If $\mathbb{K}>0$ and the semigroup $(P_t)_{t\geq 0}$ is analytic then the QMS $\mathcal{T}$ is positivity
improving.
\end{theorem}

\noindent{\it Proof.} Condition $\mathbb{K}>0$ implies that the linear dependence of operators $L_\ell$
on $a_j,a^\dagger_k$ can be inverted and so there exists complex constants $\lambda_{jk},\mu_{jk}$
($1\leq j\leq d$, $1\leq k\leq 2d$) such that
\[
a_j = \sum_{k=1}^{2d} \lambda_{jk}L_k\qquad a^\dagger_j = \sum_{k=1}^{2d}\mu_{jk}L_k.
\]
Since $\delta^{0}_G(L_\ell)=L_\ell$ for all $\ell$, for all pure state $|\psi\rangle\langle \psi|$ with $\psi\in\mathsf{h}$,
and a fixed $t>0$ the vector $P_t\psi$ belongs to $\hbox{\rm Dom}(N^n)$ for all $n\geq 0$ and, by Theorem \ref{th:GlHaNa},
the support of $\mathcal{T}_{*t}(|\psi\rangle\langle \psi|)$ contains all vectors $L_\ell P_t\psi$ with $1\leq \ell\leq 2d$.
It follows that the support of $\mathcal{T}_{*t}(|\psi\rangle\langle \psi|)$ contains all vectors
\[
a^\sharp_{j_1}\cdots a^\sharp_{j_n}P_t\psi, \qquad\qquad 1\leq j_1,\dots,j_n\leq d
\]
where $a^\sharp$ denotes either $a$ or $a^\dagger$, whence  all vectors
$G^n P_t\psi$ for $n\geq 0$.

It turns out that for any vector $v$ orthogonal to the support of $\mathcal{T}_{*t}(|\psi\rangle\langle \psi|)$
($t>0$) we have $ \langle v, G^n P_t\psi\rangle =0$ for all $n>0$ showing that the analytic function in a sector
$\left\{\, z\in\mathbb{C}-\{0\}\,\mid\, |\hbox{\rm Arg}(z)| <\theta\,\right\}$ for some
$\theta\in ]0,\pi/2[$
\[
s\mapsto \left\langle v, P_s\psi \right\rangle
\]
has zero derivatives of all orders at $t>0$ and so  $\left\langle v, P_s\psi \right\rangle=0$ for all $s>0$.
In the same way, since $P_s\psi\in\hbox{\rm Dom}(G^n)=\hbox{\rm Dom}(N^n)$ for all $n$ by Lemma \ref{lem:DGn=DNn},
$\left\langle v, a^\sharp_{j_1}\cdots a^\sharp_{j_n} P_s\psi \right\rangle=0$ for all $s>0$.
Therefore $v$ is orthogonal to the subspace generated by vectors (\ref{eq:suport3}) which turns out to be
$L_\ell$ and $P_s$ invariant contradicting irreducibility by Theorems \ref{th:subpr-invsp} and \ref{th:irreduc}.
\hfill $\square$

\subsection{The role of $H$ and an application.}
The inequality (\ref{eq:NrbG}) plays a key role in our analysis because it allows us to fix
domain problems and define operator compositions. One may note that, in the step leading to
(\ref{eq:GqbtN}), we neglected the positive term $\langle\xi, H^2 \xi\rangle$, namely
possible contribution of second order terms in $a_j,a^\dagger_k$ in the Hamiltonian $H$. \\
Let us denote by $H_0$ the sum of these terms in $H$. It may happen that, even if $\mathbb{K}$ is not
strictly positive, we have
\[
\left\langle \xi, (G^2_0+H_0^2)\xi\right\rangle \geq
\varepsilon \left\langle \xi, N^2\xi\right\rangle - k_\varepsilon \Vert\xi\Vert^2.
\]
for some $\varepsilon>0$ and so we can get our conclusion $\hbox{\rm Dom}(G)=\hbox{\rm Dom}(N)$.\\
One can also associate a matrix $\mathbb{H}$ in $M_{2d}(\mathbb{C})$ with $H_0$
\[
 \mathbb{H}= \left[ \begin{array}{cc}
 \Omega   & \kappa \\    \overline{\kappa} & \Omega^{\hbox{\scriptsize T}}
 \end{array}\right]
\]
It would be interesting to deduce the inequalities of Theorem \ref{th:DomG=DomN} from a positivity
condition on $\mathbb{K}+\mathbb{H}$ or similar but it does not seem easy in the generic case with
all parameters.

In the sequel we analyze, as a simple case in which our methods apply, the example of a system with
two bosons in a common bath of \cite{CGMZ}.
Neglecting the linear part of the Hamiltonian $H$ that plays no role as we have seen, in this
model we have $d=2$, the Kossakovski matrix and matrix $\mathbb{H}$ associated with the quadratic
part of the Hamiltonian $H$  are (as blocks of $2\times 2$ matrices)
\[
\mathbb{K}=\left[ \begin{array}{cc} \gamma^{-} & 0 \\
                                    0  & \gamma^{+} \\
                                    \end{array}\right]
\qquad
\mathbb{K}=\left[ \begin{array}{cc} \Omega & 0 \\
                                    0  & \Omega^{\hbox{\scriptsize T}} \\
                                    \end{array}\right]
\]
where $[\gamma^{\pm}_{jk}]_{1\leq j,k\leq 2}, [\Omega_{jk}]_{1\leq j,k\leq 2}$ are Hermitian matrices
with $[\gamma^{\pm}_{jk}]_{1\leq j,k\leq 2}$ positive semidefinite.
Clearly, $\kappa=0,\zeta=0$. Writing the spectral decomposition of matrices $\gamma^{\pm}$
\[
\gamma^{\pm} = \lambda_{\pm}|\varphi^{\pm}\rangle\langle \varphi^{\pm}| + \mu_{\pm}|\psi^{\pm}\rangle\langle \psi^{\pm}|
\]
where $\lambda^{\pm},\mu^{\pm}\geq 0$ and $(\varphi^{\pm}_j)_{j=1, 2}, (\psi^{\pm}_j)_{j=1, 2}$ are unit vectors
we can write generalized GKLS form of the generator with
\begin{eqnarray*}
L_1 = \lambda_{-}^{1/2} \sum_{k=1,2} \varphi_k^{-} a_k & \qquad &
L_2 = \mu_{-}^{1/2} \sum_{k=1,2} \psi_k^{-} a_k \\
L_3 = \lambda_{+}^{1/2} \sum_{k=1,2}\varphi_k^{+} a^\dagger_k & \qquad  &
L_4 = \mu_{+}^{1/2} \sum_{k=1,2}\psi_k^{+} a^\dagger_k
\end{eqnarray*}
Clearly $\mathbb{K}>0$ if and only if both matrices $\gamma^{\pm}$ are strictly positive definite i.e.
$\lambda^{\pm}>0,\mu^{\pm}>0$ and the semigroup is irreducible by Theorem \ref{th:irreduc}. \\
Unfortunately, known conditions on $G$ to generate an analytic semigroup are not immediate to check in the general case
and so we just outline two cases in which one can apply results of Section \ref{sect:pos-impr} \\
The operator $G_0$ is negative self-adjoint therefore it generates an analytic semigroup. As a consequence,
if $H$ is ``small'' (the matrix $\Omega$ is ``small'' with respect to $\gamma^{-}$ and $\gamma^{+}$, roughly speaking)
with respect to $G_0$, then by known perturbation results, the operator $G$ generates an analytic semigroup.\\
If the three matrices $\gamma^{\pm}>0$ and $\Omega$ are diagonal, then $G$ also generates an
analytic semigroup $(P_z)_{z\in \Delta}$ for $z$ in a sector
$\{\, z \in \mathbb{C}-\{0\} \, \mid\,  |\hbox{\rm Arg }(z)| < \theta \,\}$ ($\theta< \pi/2$) with $z$
depending on $\lambda^{\pm},\mu^{\pm},\Omega$ (the smaller is $\Omega$ with respect to $\gamma^{\pm}$,
the bigger the sector, i.e. the closer $\theta$ to $\pi/2$). Therefore, if $\lambda^{\pm}>0,\mu^{\pm}>0$ and $\Omega$
is ``small'' with respect to $\gamma^{-}$ and $\gamma^{+}$, also for small values of off-diagonal elements of the
three matrices $G$ generates an analytic semigroup; in these situations the Gaussian QMS is positivity improving
by Theorem \ref{th:pos-imp}.

\section{Conclusions}
\setcounter{equation}{0}
We showed that a Gaussian QMS with strictly positive Kossakowski matrix $\mathbb{K}$ is irreducible and, under the
additional assumption that the semigroup $(P_t)_{t\geq 0}$ generated by $G$ is analytic, that it is positivity improving.
Strict positivity of the $\mathbb{K}$ is, in turn, a good starting point to check that the semigroup $(P_t)_{t\geq 0}$
is analytic; however it is not a necessary condition. It would be interesting to prove rigorously that a Gaussian
QMS is positivity improving under the only condition $\mathbb{K}>0$ or, even better, under some positivity
condition involving only the matrix $\mathbb{K}$ and iterated commutators as those in Theorem \ref{th:GlHaNa}. \\
Relationships with controllability \cite{ScHer} and the analysis of some subclass, as for instance, gaussian
QMS arising from the weak-coupling limit \cite{HCQu} also deserve further investigation.

\section*{Acknowledgments}
The second author FF had the invaluable privilege to know Andrzej Kossakowski and exchange ideas
with him. His deep knowledge and far reaching insight on open quantum systems have always
been inspiring. We are comforted in the knowledge that, through his work, he has left an indelible
influence on his students, colleagues and many researchers that will keep his memory alive.

\end{document}